\def\imat  {{\rm i}}
\documentstyle[twocolumn,floats,epsf,ifthen,aps,prb]{revtex} 

\begin{document} 

 

\title{Numerical analysis of
the radio-frequency single-electron transistor operation}

\author{Valentin O. Turin 
and Alexander N. Korotkov}  
\address{ 
Department of Electrical Engineering, University of California, 
Riverside, CA 92521-0204. 
} 
\date{\today} 
 
\maketitle 
 
\begin{abstract} 
We have analyzed numerically the response and noise-limited charge 
sensitivity of a radio-frequency single-electron transistor (RF-SET)
in a non-superconducting state using the orthodox theory. In particular, 
we have studied the performance dependence on the quality factor $Q$ 
of the tank circuit  
for $Q$ both below and above the value corresponding to the impedance 
matching between the coaxial cable and SET. 
\end{abstract} 
 
\narrowtext 
 
\vspace{0.0cm} 

\section{Introduction} 

        An important drawback of the conventional single-electron
transistor \cite{Av-Likh,Fulton} (SET) is its relatively large output 
resistance which should be much larger than the quantum resistance
$R_Q=h/4e^2\simeq 6.5\,\mbox{k}\Omega$. This limits the operation frequency 
of the prospective integrated single-electron circuits
\cite{Likh-rev,Kor-rev} and imposes a severe frequency limitation 
for individual SETs used nowadays as electrometers. Estimating the 
total capacitance of a wire delivering SET signal from a cryostat to 
outside electronics as 1 nF, we get the time constant on the order 
of $1 \, \mbox{nF}\times 10^5 \, \Omega = 10^{-4}\, \mbox{s}$; therefore the
operating frequency is limited to few kHz, which is a typical value
achievable by conventional SET setups.\cite{Starmark} 
        While the operating frequency can be significantly increased  
by placing a preamplifier in a close vicinity of the SET,  
\cite{Pettersson,Visscher}
the preferred at present solution of the problem is the use of the 
radio-frequency SET \cite{Schoelkopf} (RF-SET) which in many instances 
has already replaced the traditional SET setup. 

        The principle of the RF-SET operation is somewhat similar to 
the operation of the radio-frequency superconducting quantum 
interference device \cite{Clarke} (RF-SQUID) and is based on the 
microwave reflection 
\cite{Schoelkopf,Wahlgren,Aassime,Aassime-2,Aassime-3,Stevenson,Segall,%
Lehnert,Bladh,Buehler-1,Buehler-2} 
from a tank (LC) circuit containing the SET (Fig.\ \ref{fig1}); 
another possibility is to use the transmitted wave. \cite{Fujisawa,Cheong}
The measured SET input signal changes effective SET resistance and 
affects the intensity and the phase of the reflected (or transmitted) wave
that is later sensed by either homodyne detection or simple rectification 
(to separate incoming and reflected waves a directional coupler can be used).
The high operation frequency of the RF-SET 
is due to the signal propagation by the microwave, so that the SET does not
need to charge the whole output wire, while the tank circuit tuned 
in resonance works as an impedance transformer providing a better match 
between the effective SET differential resistance 
$R_d$ ($\sim 10^5\, \Omega$) and the microwave cable wave impedance $R_0$ 
(typically $50\, \Omega$). 

For a good matching the ``unloaded'' quality factor 
$Q\equiv \sqrt{(L_T/C_T)}/R_0$ of the tank circuit consisting of inductance
$L_T$ and capacitance $C_T$ should be comparable to $\sqrt{R_d/R_0} \sim 50$.
While a much lower value, $Q=6$, was used in the first experiment,
\cite{Schoelkopf} the values close to the matching condition (sometimes
even higher) are typically 
used at present. Increase of the $Q$-factor obviously decreases the RF-SET 
bandwidth limited by $\sim \omega /2Q_L$ where 
$\omega \approx 1/\sqrt{L_TC_T}$ 
is the ``carrier'' microwave frequency and $Q_L$ is the ``loaded'' 
$Q$-factor, which also takes into account the effect of the SET 
(see below). 
 The straightforward design \cite{Schoelkopf} shown in Fig.\ \ref{fig1} 
can be somewhat modified to reduce the bandwidth further 
for a dense multiplexing;\cite{Stevenson} however, in this paper we will
consider only the original design. 

        The RF-SET bandwidth as wide as 100 MHz has been demonstrated 
\cite{Schoelkopf} using a relatively high carrier frequency $\omega /2\pi
=1.7\, \mbox{GHz}$ and relatively low $Q$-factor $Q=6$. However, to improve
experimental RF-SET sensitivity, it happens to be advantageous to reduce 
the carrier 
frequency by few times (to reduce the noise contribution from the  
amplifier) and also increase $Q$ (closer to the impedance matching regime),
so that a practical bandwidth at present is about 10 MHz (for example, 
the bandwidth of 7 MHz for the carrier frequency of 332 MHz has been reported
in Ref.\ \cite{Aassime-2}). 

        The high operation frequency of the RF-SET makes it easily possible 
to avoid the 1/$f$ noise limitation of the SET sensitivity which is typically
dominant at frequencies $f\lesssim 10^4\, \mbox{Hz}$, and so work in the 
region of the shot noise limited sensitivity.\cite{Kor-92,Kor-94} 
Though experimentally the contribution from the amplifier noise is still 
comparable or larger than the SET noise, a relatively rapid improvement of 
the RF-SET charge sensitivity from $1.2\times 10^{-5}\, e/\sqrt{\mbox{Hz}}$
at 1.1 MHz in the first experiment \cite{Schoelkopf} to the value 
$3.2\times 10^{-6}\, e/\sqrt{\mbox{Hz}}$ (4.8 $\hbar$ in energy units) 
at 2 MHz 
reported in Ref.\ \cite{Aassime-2} assures that the pure shot noise limit 
will be achieved pretty soon. (The above numbers can be compared with 
the sensitivity $2\times 10^{-5}\, e/\sqrt{\mbox{Hz}}$ at 4.4 kHz of a
purely conventional SET reported in Ref.\ \cite{Starmark}, the value 
$8\times 10^{-6}\, e/\sqrt{\mbox{Hz}}$ at 10 Hz for the ``stacked''
SET, \cite{Krupenin} and the sensitivity $6\times 10^{-6}\, 
e/\sqrt{\mbox{Hz}}$ at 45 Hz for the SET made of carbon 
nanotubes. \cite{Roschier}) 

        One of important potential applications of the RF-SET is for 
the readout of the charge qubits in a solid-state quantum computer. 
\cite{Aassime,Bladh,Lehnert,Buehler-1,Buehler-2,Echternach} 
The possibility of a single-shot qubit measurement requires 
fast enough distinguishing between two charge states to avoid significant
qubit evolution during measurement. This requires sufficiently wide  
RF-SET bandwidth and most importantly good enough charge sensitivity.
The estimates \cite{Aassime} show that the single-shot qubit measurement
is almost within the reach of present-day RF-SET performance; however, 
reliable measurement still requires significant improvement in sensitivity. 
This makes very important the question of ultimate (theoretical) 
RF-SET sensitivity. The ultimate RF-SET sensitivity is also a crucial 
parameter for monitoring quantum dynamics of nanomechanical resonators
\cite{Armour,Irish,Knobel-Nature} and for a variety of RF-SET applications 
as an electrometer in classical single-electron devices. 

        In spite of significant experimental activity on RF-SETs,
we are aware of only few theoretical papers on the RF-SETs. 
The basic theory of the shot noise limited charge sensitivity of the RF-SET 
has been developed in Ref.\ \cite{Kor-RF}. A similar theory has been applied
to the sensitivity analysis for the RF-SET-based micromechanical displacement
detector. \cite{Blencowe,Zhang,Zhang-2} Some theoretical analysis of
the transmission-type RF-SET can be found in Ref.\ \cite{Cheong}. 
The theory of a somewhat related device, radio-frequency Bloch-transistor,
has been developed in Ref.\ \cite{Zorin}. 
In our opinion, the RF-SET definitely requires further theoretical 
attention, since many questions about RF-SET performance has not yet been
answered theoretically.

        In this paper we extend the theory of Ref.\ \cite{Kor-RF}
to the case of arbitrary $Q$-factor of the tank circuit, removing the 
assumption (strongly violated in the present-day experiments) 
of $Q$ being much smaller than the impedance matching value.
We calculate the response and sensitivity of the normal-metal RF-SET 
and optimize these magnitudes numerically over the rf wave amplitude 
and the SET background charge. 
Then we study the dependence 
of the optimized RF-SET response and sensitivity on the tank $Q$-factor, 
operation temperature, SET resistance, carrier frequency, and SET asymmetry 
due to asymmetric biasing. Some results of this paper have been presented
earlier in a short form. \cite{Turin-Kor}

        \section{Model and calculation methods}

\begin{figure}
\centerline{ 
\epsfxsize=3.3in
\epsfbox{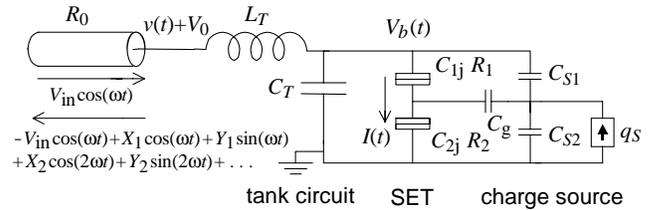} 
}
\vspace{0.3cm}
\caption{Schematic of the RF-SET. The current $I(t)$ through the SET
(two tunnel junctions with capacitances $C_{1j}$ and $C_{2j}$)
 affects the quality factor of the tank circuit ($L_T$ and $C_T$) and
therefore the amplitude and phase of the reflected rf wave propagating 
along coaxial cable ($R_0$). The change of the source charge $q_S$ changes 
the effective SET background charge $q_0$ and can be monitored 
via measurement of the reflected rf wave.}
\label{fig1}\end{figure}

        The schematic of the RF-SET used in our analysis is shown
in Fig.\ \ref{fig1}. The SET consists of two tunnel junctions with 
capacitances $C_{1j}$ and $C_{2j}$ and resistances $R_1$ and $R_2$. 
The SET is coupled via gate capacitance $C_g$ to the measured charge 
source (for example, a single-electron box or a similar structure), 
which is characterized by the charge $q_S$ and capacitances 
$C_{S1}$ and $C_{S2}$ to the SET leads (the total source capacitance 
is $C_S=C_{S1}+C_{S2}$). Assuming constant $q_S$ (neglecting backaction
from the SET), it is easy to show  that the SET coupled to the 
charge source is equivalent to the simple double-junction SET structure
with parameters 
        \begin{eqnarray}
&&  C_1= C_{1j} + \frac{C_gC_{S1}}{C_g+C_S} , 
        \label{C1}\\
&&  C_2= C_{2j} + \frac{C_gC_{S2}}{C_g+C_S} ,
        \label{C2}\\
&& q_0=q_{00}+ q_S \frac{C_g}{C_g+C_S}, 
        \label{q0}\end{eqnarray} 
where $C_1$ and $C_2$ are effective junction capacitances (the total
SET island capacitance is $C_{\Sigma}=C_1+C_2$), and 
the total induced charge of the SET island $q_0$ is the sum of the
initial background charge $q_{00}$ and the contribution
from the measured charge source.
      For numerical results we have used the ``orthodox'' model
\cite{Av-Likh} for a normal-metal SET (see Appendix). 

\subsubsection*{Linear analysis of the reflected wave} 

        The current $I(t)$ through the SET affects the quality factor
of the tank circuit consisting of inductance $L_T$ and capacitance $C_T$,
while the contribution to the tank circuit capacitance is neglected,
assuming $C_T \gg \max [C_{1j},C_{2j},C_{S1}C_{S2}/(C_{S1}+C_{S2})]$.
(This condition is usually well satisfied experimentally; it also allows us
to neglect the effect of single electron jumps on the tank circuit 
oscillations.) The quality factor of the tank circuit is also affected 
by the wave impedance $R_0$ of the cable. For a simple linear analysis
let us substitute the SET by an effective differential resistance $R_d$.
Then, assuming the case of weak damping, we can simply add the contributions
from $R_0$ and $R_d$, so that the total (``loaded'') quality factor
of the tank circuit is
        \begin{eqnarray}
&& Q_L =(1/Q + 1/Q_{SET})^{-1},
        \\
&& Q = \frac{\sqrt{L_T/C_T}}{R_0}, \,\,\, 
        Q_{SET} = \frac{R_d}{\sqrt{L_T/C_T}},
        \end{eqnarray}
where the ``unloaded'' quality factor $Q$ corresponds to absence of the
SET, $R_d=\infty$, while $Q_{SET}$ corresponds to the damping by the SET
only, $R_0=0$. Notice that $Q$ is fixed by the RF-SET design, while
$Q_{SET}$ and therefore $Q_L$ depend on the operating conditions;
so even though $Q_L$ is a more physically meaningful quantity than $Q$, 
in this paper we consider unloaded $Q$ as an independent parameter
and call it as $Q$-factor. 

        For the incoming voltage wave $\hat{V}_{in}\exp (\imat \omega t)$, 
the reflected wave $\alpha \hat{V}_{in}\exp (\imat \omega t)$ is determined 
by the complex reflection coefficient $\alpha$: 
        \begin{equation}
\alpha = \frac{Z-R_0}{Z+R_0}, \,\,\,\, Z=\imat\omega L_T + 
        \frac{1}{\imat\omega C_T +1/R_d} . 
        \end{equation}
Since the measured charge signal changes the SET resistance $R_d$, the 
RF-SET response is proportional to $d(\alpha\hat{V}_{in})/dR_d$.
Using the first order approximation close to the resonant frequency
$\omega_0 = 1/\sqrt{L_TC_T}$ of the tank circuit: 
\begin{equation}
Z \approx \frac{L_T/C_T}{R_d} + 2\,\imat\sqrt{L_T/C_T}\,\frac{\Delta\omega}
{\omega_0}, \,\,\, \Delta \omega =\omega -\omega_0 
\end{equation}
(in particular, this approximation neglects the  shift of
the resonant frequency, which is a second-order effect) it is easy to get 
        \begin{equation}
\frac{d\alpha}{dR_d} \approx \frac{-2R_0}{R_d^2}\, 
\frac{Q^2}{(1+\frac{Q^2}{R_d/R_0})^2}\,
\frac{1}{(1+2\imat Q_L\Delta\omega /\omega_0)^2}. 
        \label{dadr}\end{equation} 
For fixed values of $R_0$ and $R_d$, this expression increases with 
$Q$ for small $Q$, thus showing the need of $Q$-factor increase 
to achieve a better RF-SET response. The maximum is reached at  
        \begin{equation}
        Q=\sqrt{R_d/R_0},
        \end{equation}
which is the case of practically matched impedances, $Z\approx R_0$, 
at frequencies close to resonance,   
and also corresponds to the condition  $Q=Q_{SET}=2Q_L$. Notice that
in this case the reflection practically vanishes, $\alpha \approx 0$. 

        Actually, Eq.\ (\ref{dadr}) is not really relevant, because
the rf amplitude $V_{in}$ should depend on $Q$ in order to maintain
approximately constant amplitude $|\hat{V}_b|$ of the SET bias voltage 
oscillations, which is determined by the voltage scale of the Coulomb
blockade. Taking into account the relation
        \begin{equation}
\hat{V}_{in}/\hat{V}_b=(Z+R_0)(\imat \omega C_T +1/R_d)/2,  
        \label{VinVb}\end{equation}
we obtain 
        \begin{equation}
\frac{d\alpha}{dR_d} \left| \frac{\hat{V}_{in}}{\hat{V}_b}\right| 
\approx \frac{- R_0}{R_d^2}\, 
\frac{Q}{1+\frac{Q^2}{R_d/R_0}}\,
\frac{|1+2\imat Q_L\Delta\omega /\omega_0|}
{(1+2\imat Q_L\Delta\omega /\omega_0)^2},
        \label{resp-lin}\end{equation}
which is somewhat similar to Eq.\ (\ref{dadr}) and also reaches 
maximum at $Q=\sqrt{R_d/R_0}$. Notice that this condition optimizes 
only the RF-SET response, while the shot-noise-limited sensitivity 
can (and as will be seen later does) have completely different
dependence on $Q$. 

\subsubsection*{Full analysis of the reflected wave} 

        The linear analysis discussed above can be used only as an 
estimate because of the significant nonlinearity of the SET current-voltage
($I-V$) dependence. For a more exact analysis \cite{Kor-RF} we use 
the Kirchhoff's rules 
taking into account the current $I(t)$ through the SET. Let us separate the
voltage $V_0+v(t)$ at the end of the rf cable into the dc component
$V_0$ (which can be supplied via the bias-tee \cite{Schoelkopf}) and
rf component $v(t)=V_{in}(t)+V_{out}(t)$, where 
$V_{in}(t)=V_{in}\cos{\omega t}$ is the incoming wave (from now on
we do not use complex representation) and $V_{out}(t)$ is the outgoing
reflected wave. The differential equation for the rf component $v(t)$ is 
        \begin{eqnarray}
&& \ddot{v}/\omega_0^2 +\dot{v}/Q\omega_0 + v = 
2(1-\omega^2 /\omega_0^2)V_{in}\cos{\omega t}
        \nonumber \\ 
&& \hspace{3.2cm} -R_0 [I(t)-\langle I\rangle ],
        \label{difeq}\end{eqnarray}
where $\langle I\rangle$ is the current through the SET averaged over time
much longer than $\omega^{-1}$, and the time dependence of the SET current
$I(t)$ can be found self-consistently from the time dependence 
of its bias voltage 
        \begin{equation}
V_b(t) = V_0 + v (t) + [2V_{in}\omega \sin{\omega t} +
\dot{v}(t)]\, Q/\omega_0 .
        \label{Vb}\end{equation}
In this paper we assume 
that the rf frequency is small compared to the frequency of electron
tunneling through the SET, $\omega \ll I/e$, so that the SET shot noise 
is a small contribution compared to the deterministic part of the 
SET current $I(t)$, which is calculated using dc $I-V$ curve and $V_b(t)$
(actually, it is still OK if this condition is not satisfied during 
some fraction of the period due to Coulomb blockade, since the small
current does not affect the oscillations significantly).

In a steady state (assuming that the induced SET charge $q_0$ does
not change with time) the reflected wave can only contain the incoming 
frequency $\omega$ and its overtones:
        \begin{equation}
V_{out}(t) = -V_{in} \cos{\omega t} +\sum_{n=1}^\infty
[ X_n \cos{n\omega t} +Y_n \sin{n\omega t} ] .
        \label{Vout}\end{equation}
We separate the term $-V_{in}\cos{\omega t}$ mainly to follow the notations 
of Ref.\ \cite{Kor-RF}, even though it really makes sense only for a low-$Q$ 
case at $\omega \approx \omega_0$, when all $X_n$ and $Y_n$ are small.
Using the substitution 
$v(t)=\sum_{n=1}^\infty [X_n \cos{n\omega t} +Y_n \sin{n\omega t} ]$
in Eq.\ (\ref{difeq}), we find  
the coefficients $X_n$ and $Y_n$ as 
        \begin{eqnarray}
&& X_n=R_0 Q \, \frac{n\tilde{\omega}\,a_n -
Q (1-n^2\tilde{\omega}^2)\, b_n 
} 
{n^2\tilde{\omega}^2 + Q^2(1 -n^2\tilde{\omega}^2)^2}  
\nonumber\\
&& \hspace{0.9cm} + \frac{2 Q^2 (1-\tilde{\omega} ^2)^2} 
{\tilde{\omega}^2 +Q^2(1-\tilde{\omega}^2)^2 }
\, V_{in}\delta_{1n} \, , 
\label{Xn}
        \\ 
&& 
 Y_n=- R_0 Q \, \frac{n\tilde{\omega} \, b_n +
Q (1-n^2\tilde{\omega}^2)\, a_n 
}
{n^2\tilde{\omega}^2+Q^2(1-n^2\tilde{\omega}^2)^2} 
        \nonumber\\
&& \hspace{0.9cm} 
+  
\frac{2 Q \tilde{\omega}  
(1-\tilde{\omega}^2)}
{ \tilde{\omega}^2+ Q^2(1-\tilde{\omega}^2)^2}\, 
V_{in}\delta_{1n}\, ,
        \label{Yn}\end{eqnarray}
where
\begin{equation}
a_n =2 \langle I(t)\sin{n\omega t} \rangle , \,\,\,
b_n =2 \langle I(t)\cos{n\omega t} \rangle  
\label{anbn}\end{equation}
(averaging is over the oscillation period), 
 $\tilde{\omega}\equiv \omega /\omega_0$ is the normalized frequency,
$\delta_{1n}$ is the Kronecker symbol, 
and
the current $I(t)$ is calculated self-consistently using the SET
bias voltage
        \begin{eqnarray} 
&& V_b(t) = V_0 + 2Q\frac{\omega}{\omega_0} V_{in} \sin{\omega t}
+\sum_{n=1}^\infty \left[ (X_n+\frac{Qn\omega}{\omega_0}Y_n) \right. 
        \nonumber \\
&& \hspace{1.2cm} 
\left. \times  \cos{n\omega t}
+ (Y_n-\frac{Qn\omega}{\omega_0}X_n) \sin{n\omega t} \right] .
        \label{Vbiter}\end{eqnarray}

        Because of the resonant behavior of the tank circuit at $Q \gg 1$, 
the contribution of overtones ($n\geq 2$) in the reflected signal is 
always small if $\omega \approx \omega_0$ 
(since the overtones are far from resonance).  This can be easily seen 
from Eqs.\ (\ref{Xn})--(\ref{Yn}), especially in the case $\omega =\omega_0$,
when they are significantly simplified: 
        \begin{eqnarray}
&& X_n = R_0 Q \,\frac{na_n-Q(1-n^2)b_n}{n^2+Q^2(1-n^2)^2} \, , 
        \\
&& Y_n = - R_0 Q \,\frac{nb_n+Q(1-n^2)a_n}{n^2+Q^2(1-n^2)^2} \, .
        \end{eqnarray}

        Therefore, a linear (one-frequency) approximation 
in which only $X_1$ and $Y_1$ 
are taken into account, works very well in this case. In our numerical 
analysis we used the result of the linear approximation as a starting
point of the iterative procedure [$V_b(t) \rightarrow (a_n,b_n)\rightarrow
(X_n,Y_n)\rightarrow V_b(t) \rightarrow \dots$] to solve selfconsistently 
Eqs.\ (\ref{Xn})--(\ref{Vbiter}) taking into account few (typically 
3-5) overtones. We have checked numerically that the account of overtones 
typically gives a small correction in the case of a reasonably large 
$Q$-factor and $\omega \approx \omega_0$. 
 
        In the linear approximation the SET bias voltage has only 
one frequency component: $V_b(t) =V_0+A_b\sin (\omega t+\phi)$ and
therefore for the calculation of $a_1$ and $b_1$ [see Eq.\ (\ref{anbn})] 
the SET can be simply 
replaced by the effective resistance 
        \begin{equation} 
R_d = \frac{\pi A_b}{\int_0^{2\pi} I(V_0+A_b\sin{x}) \sin{x} \, dx} \, ,
        \end{equation} 
where $I(V)$ is the SET current-voltage dependence.
[Notice that $\int_0^{2\pi} I(V_0+A_b\sin{x}) \cos{x} \, dx =0$, 
i.e.\ there is no effective reactance contribution.] 

        Hence, this linear (one-frequency) approximation is 
completely equivalent to the case of a resistor instead of the SET,   
considered in the previous subsection. The only new condition
is a selfconsistent relation between the effective resistance $R_d$ and
the amplitude $A_b$ of the SET bias voltage. The amplitude $A_b$ (which 
depends on $R_d$) can be
calculated either using Eq.\ (\ref{VinVb}) in which $A_b=|\hat{V}_b|$ 
or using Eq.\ (\ref{Vbiter}), which gives
        \begin{equation}
A_b= \sqrt{[Q\tilde{\omega}(2V_{in}-X_1)+Y_1]^2
+(X_1+Q\tilde{\omega}Y_1)^2 } ,
        \end{equation}
while the components $X_1$ and $Y_1$ are given by equations 
        \begin{eqnarray}
&& X_1= \frac{2Q^2}{Q^2+\tilde{R}} 
\frac{\tilde{\omega}^2 +(1-\tilde\omega^2) \tilde{R} 
\frac{1+(1-\tilde\omega^2)\tilde{R}}{Q^2+\tilde{R}}}
{\tilde\omega^2+\frac{Q^2[1+(1-\tilde\omega^2)\tilde{R}]^2}
{(Q^2+\tilde{R})^2}} \,  V_{in}, 
        \label{X1}\\
&& Y_1= \frac{2\tilde\omega Q}{Q^2+\tilde{R}} 
\frac{-\frac{Q^2[1+(1-\tilde\omega^2)\tilde{R}]}{Q^2+\tilde{R}}
+(1-\tilde\omega^2)\tilde{R}} 
{\tilde\omega^2+\frac{Q^2[1+(1-\tilde\omega^2)\tilde{R}]^2}
{(Q^2+\tilde{R})^2}} \,  V_{in}, 
        \label{X2}\end{eqnarray} 
where $\tilde{R}\equiv R_d/R_0 \gg 1$ (these equations can obviously be 
rewritten in a shorter way; however, they become less transparent 
to analyze). 

        The equations (\ref{X1})--(\ref{X2}) significantly simplify 
in the case $\omega =\omega_0$:
        \begin{eqnarray}
&& X_1=\frac{2Q^2 (Q^2+\tilde{R})}{(Q^2+\tilde{R})^2+Q^2} \, V_{in},
        \label{X1s}\\
&& Y_1=\frac{-2Q^3}{(Q^2+\tilde{R})^2+Q^2} \, V_{in} , 
        \label{Y1s}\end{eqnarray}
from which it is clear that $|Y_1/X_1|\ll 1$ for $Q\gg 1$ 
, and therefore 
        \begin{equation}
V_b(t) \approx V_0 + \frac{2 Q\tilde{R}V_{in}}{Q^2+\tilde{R}}\sin{\omega t}=
V_0 + 2 Q_L V_{in} \sin{\omega t}.
        \label{Vbsimple}\end{equation}

\subsubsection*{Response and noise-limited sensitivity} 

        So far we have implicitly assumed that the SET current $I(t)$ 
depends on time only because of the periodic time dependence of the 
SET bias voltage $V_b(t)$. However, $I(t)$ has also a small noise component,
the magnitude of which depends on the bias voltage and therefore also has 
a periodic time dependence. The shot  noise of the SET current 
leads to the 
fluctuations of the parameters $a_n$ and $b_n$ defined by Eq.\ (\ref{anbn})
and consequently to the fluctuations of the reflected wave quadratures 
$X_n$ and $Y_n$. Since the noise of $X_n$ and $Y_n$ can be meaningfully 
discussed only at frequencies less than $\omega /Q_L$, which is much less
than the typical frequency $I/e$ of electron tunneling in the SET, 
it is sufficient to consider the low-frequency limit of the SET shot noise. 

        The low-frequency spectral densities of $a_n$ and $b_n$, 
and their mutual spectral density can be calculated as 
        \begin{eqnarray}
&&      S_{an}=4\langle S_I(t) \sin^2 n\omega t\rangle , \,\,\,\,
        S_{bn}=4\langle S_I(t) \cos^2 n\omega t\rangle , 
        \\
&&      S_{an,bn}= 2 \langle S_I(t) \sin 2 n\omega t\rangle , 
        \end{eqnarray}
where the averaging is over the oscillation period, 
 $S_I(t)$ is the low- (zero-) frequency spectral density of the SET 
current (see Appendix),  
and the time dependence comes from oscillating bias voltage $V_b(t)$.
Consequently, the low-frequency spectral densities of $X_n$ and $Y_n$
fluctuations, and their mutual spectral density are 
        \begin{eqnarray}
&&      S_{Xn}=c_{n}^2 \langle S_I(t) \sin^2 n\omega t\rangle + 
              d_{n}^2 \langle S_I(t) \cos^2 n\omega t\rangle 
        \nonumber \\ 
&& \hspace{1cm} - c_{n}d_{n} \langle S_I(t) \sin 2n\omega t\rangle, 
        \\
&&      S_{Yn}=d_{n}^2 \langle S_I(t) \sin^2 n\omega t\rangle + 
              c_{n}^2 \langle S_I(t) \cos^2 n\omega t\rangle 
        \nonumber \\ 
&& \hspace{1cm} +c_{n}d_{n} \langle S_I(t) \sin 2n\omega t\rangle, 
        \\
&&        S_{Xn,Yn} = c_{n} d_{n} \langle S_I(t) \cos 2n\omega t\rangle 
        \nonumber \\
&& \hspace{1cm} 
  +\frac{1}{2}(d_{n}^2-c_{n}^2)\langle S_I(t)\sin 2n\omega t\rangle , 
        \end{eqnarray} 
where
        \begin{eqnarray}
&& c_{n}= \frac{2 R_0 Q n \tilde\omega}
{n^2\tilde{\omega}^2 + Q^2(1 -n^2\tilde{\omega}^2)^2} ,
        \label{f1n}\\
&& d_{n}= \frac{2 R_0 Q^2 (1-n^2 \tilde\omega^2)}
{n^2\tilde{\omega}^2 + Q^2(1 -n^2\tilde{\omega}^2)^2} .
        \label{f2n}\end{eqnarray}
Notice much simpler equations for $S_{X_1}$ and $S_{Y_1}$
in the case $\omega =\omega_0$: 
        \begin{eqnarray}
&& S_{X_1} = 4R_0^2Q^2 \langle S_I(t) \sin^2 \omega t\rangle ,
        \label{SX1}\\
&& S_{Y_1} = 4R_0^2Q^2 \langle S_I(t) \cos^2 \omega t\rangle .
        \label{SY1}\end{eqnarray}

        The change of the measured charge $q_S$ can in principle be
monitored via the change of any quadrature $X_n$ or $Y_n$. The corresponding
{\it responses} can be defined as the derivatives $dX_n/dq_S$ or $dY_n/dq_S$, 
while the corresponding {\it sensitivities} (minimal detectable charge 
$\delta q_S$ for a small measurement bandwidth $\Delta f$) are 
        \begin{equation}
\delta q_{S,Xn} = \frac{\sqrt{S_{Xn}\Delta f}}{|dX_n/dq_S|}, \,\,\, 
\delta q_{S,Yn} = \frac{\sqrt{S_{Yn}\Delta f}}{|dY_n/dq_S|} . 
        \end{equation}

Numerical calculations show that in the usual case $\omega \approx \omega_0$ 
thus defined RF-SET sensitivity for 
overtones ($n\geq 2$) can be comparable to the sensitivity using
the carrier frequency ($n=1$). 
However, because of relatively small amplitude of reflected overtones 
in this case, 
their monitoring is impractical, 
and so we mainly consider monitoring of $X_1$ and 
$Y_1$ which are referred below as $X$ and $Y$.

        Monitoring both quadratures $X$ and $Y$ simultaneously, 
one can improve the sensitivity compared with monitoring of only 
one quadrature. It is easy to show that 
the resulting sensitivity can be obtained as the optimization over
angle $\varphi$ of the sensitivity corresponding to monitoring the 
linear combination $X^\ast =X\cos \varphi +Y\sin \varphi$ (experimentally, 
this is just a phase shift in the homodyne detector; 
notice that the contribution $-V_{in} \cos 
\omega t$ is noiseless by assumption). The optimum sensitivity is achieved at
$\tan \varphi = (S_X dY/dq_S - S_{XY} dX/dq_S)/(S_Y dX/dq_S -S_{XY} dY/dq_S)$
and the resulting sensitivity is
\begin{equation} 
\frac{\delta q_{S,X^\ast}}{\sqrt{\Delta f}} = \left[
\frac{S_XS_Y-S_{XY}^2}
{S_Y(\frac{dX}{dq_S})^2+S_X(\frac{dY}{dq_S})^2-2S_{XY}
\frac{dX}{dq_S}\frac{dY}{dq_S}} \right]^{1/2} ,
\label{dq1}\end{equation}
that can be rewritten as \cite{Kor-RF} 
        \begin{equation}
\frac{\delta q_{S,X^\ast}}{\sqrt{\Delta f}} = \left[
\frac{1-K^2}{(\delta q_{S,X})^{-2} +(\delta q_{S,Y})^{-2}
-2K/\delta q_{S,X}\delta q_{S,Y}} \right]^{1/2},
        \label{dq2}\end{equation}
where $K\equiv S_{XY}/\sqrt{S_XS_Y}\, \mbox{sign} [(dX/dq_S)(dY/dq_S)]$ 
is the noise correlation factor 
[$S_{XY}$ is a real magnitude because we consider only low-frequency
sensitivity; for finite frequency $S_{XY}$ in Eq.\ (\ref{dq1}) should be 
replaced by $\mbox{Re} S_{XY}$]. 
        Notice that the response optimization for $X^\ast$ monitoring
is achieved at different phase shift: $\tan \varphi =(dY/dq_S)/ 
(dX/dq_S)$ that leads to the response $dX^\ast /dq_S = 
[(dX/dq_S)^2+(dY/dq_S)^2]^{1/2}$.

        Let us also consider the case when the reflected wave is monitored
by simple rectification. Assuming for simplicity the monitoring of only the 
first harmonic amplitude $A_1=[(X-V_{in})^2+Y^2]^{1/2}$ (overtones 
are filtered out), the RF-SET sensitivity can be calculated as 
        \begin{equation}
\frac{\delta q_{S,A1}}{\sqrt{\Delta f}} = 
\frac{[(X-V_{in})^2S_X+Y^2S_Y+2(X-V_{in})YS_{XY}]^{1/2}}
{|(X-V_{in}) dX/dq_S +Y dY/dq_S|} 
        \label{dq3}\end{equation}
while the response is obviously $dA_1/dq_S=[(X-V_{in})dX/dq_S+YdY/dq_S]/A_1$.
The formulas for monitoring of $n$th overtone are similar, 
except $V_{in}$ does not contribute to the amplitude $A_n$.

        In the most interesting for practice case $\omega = \omega_0$ 
the magnitude of $Y$ is small in comparison with $X$ for $Q\gg1$ [see Eqs.\ 
(\ref{X1s})--(\ref{Y1s})], and the correlation factor $K$ also 
vanishes [because $d_1=0$ and the SET bias voltage contains mostly 
$\sin$-component -- see Eq.\ (\ref{Vbsimple})]. 
Then Eq.\ (\ref{dq2}) for the best 
homodyne detection practically coincides with Eq.\ (\ref{dq3}) 
for rectification 
and reduces to the sensitivity $\delta q_{S,X}/\sqrt{\Delta f} =
\sqrt{S_X}/(dX/dq_S)$ for monitoring of $X$ component only (similarly,
the formulas for response also practically coincide). 
 Because of that, the numerical results for RF-SET response and sensitivity
in the case $\omega =\omega_0$ will assume monitoring of $X$ quadrature.

\vspace{0.1cm}

        Since the increments of the measured charge $q_S$ and 
the effective SET charge $q_0$ are related by a constant factor,
        \begin{equation}
\delta q_S = \delta q_0 (1+C_S/C_g),
        \end{equation}
the RF-SET response and sensitivity in respect to $q_S$ and in respect
to $q_0$ differ by the factor $1+C_S/C_g$. 
All our numerical results will be in terms of $q_0$ measurement,
so for $X$ monitoring we will use the derivative 
$dX/dq_0$ as a measure of the RF-SET response and the magnitude 
        \begin{equation}
     \frac{\delta q_0}{\sqrt{\Delta f}}= \frac{\sqrt{S_{X}}}{|dX/dq_0|}
        \label{dq0X}\end{equation}
as a measure of the sensitivity.

        \section{Numerical results}

        We have studied numerically the dependence of the RF-SET
response and sensitivity on various parameters, which include  
``fixed'' parameters (which cannot be easily changed in an experiment) 
and the parameters 
of the operation point. The fixed parameters are: effective junction 
capacitances of the SET $C_1$ and $C_2$ (we assume $C_1=C_2$ unless
mentioned otherwise), resistances $R_1$ and $R_2$ 
(we always assume $R_1=R_2$), 
temperature $T$, cable wave impedance $R_0$ (we always assume 
$R_0=50\,\Omega$), the tank circuit frequency 
$\omega_0=1/\sqrt{L_T C_T}$, and the $Q$-factor $Q=\sqrt{L_T/C_T}/R_0$.
The operating point parameters are the effective SET background charge
$q_0$, dc bias voltage $V_0$,
amplitude $V_{in}$ of the incident wave, and its frequency $\omega$
(in most cases we assume $\omega =\omega_0$). 

We use the SET parameters for normalization, so that a natural unit
for temperature is $e^2/C_\Sigma$ (where $C_\Sigma =C_1+C_2$), 
the voltage unit is $e/C_\Sigma$, the RF-SET response $dX/dq_0$ (or 
$dY/dq_0$) can be measured in units $1/C_\Sigma$, and the unit for 
sensitivity $\delta q_0 / \sqrt{\Delta f}$ is $e(R_\Sigma C_\Sigma)^{1/2}$
(where $R_\Sigma =R_1+R_2$). Notice that all considered magnitudes  
have a simple scaling with $C_\Sigma$; however, there is no simple 
scaling with $R_\Sigma$ because of the dimensionless parameter 
$R_\Sigma /R_0$.

\subsubsection*{Operation point optimization} 

\begin{figure}
\centerline{ 
\epsfxsize=3.0in
\epsfbox{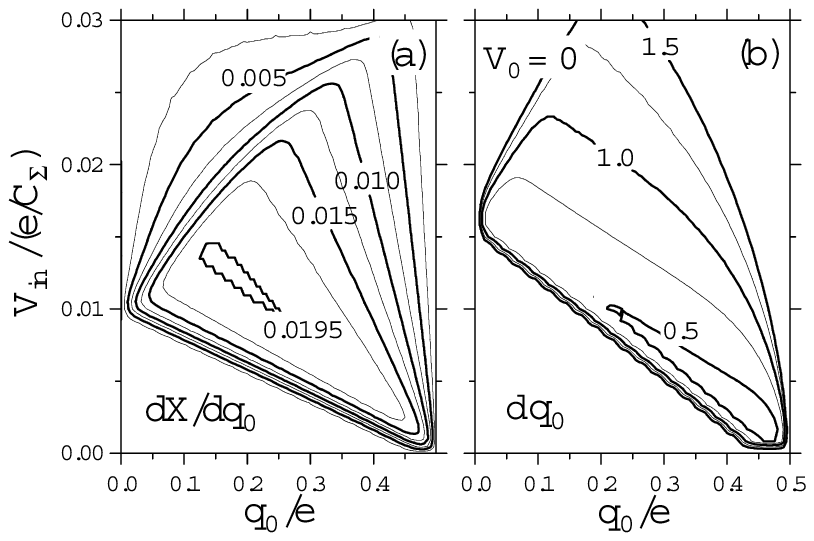} 
}
\centerline{ 
\epsfxsize=3.0in
\epsfbox{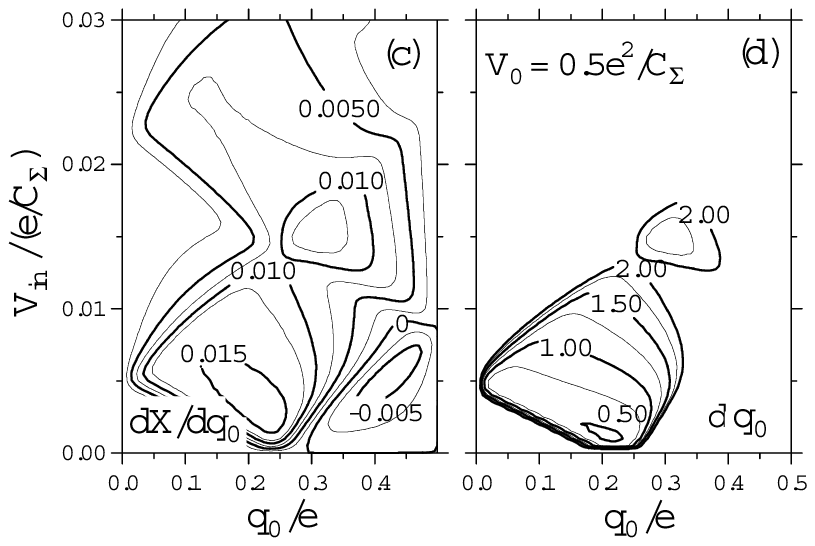} 
}
\vspace{0.2cm}
\caption{Contour plots of  (a, c) the RF-SET response $dX/dq_0$ 
(in units $C_\Sigma^{-1}$)  and 
(b, d) the noise-limited sensitivity $\delta q_0/\sqrt{\Delta f}$ (in units 
$e\sqrt{R_\Sigma C_\Sigma}$) on the plane of the SET background charge
$q_0$ and the amplitude $V_{in}$ of incoming rf wave for $Q=50$,  
$T=0.01\, e^2/C_\Sigma$, $R_\Sigma /R_0 =2000$, and $\omega =\omega_0$.
$V_0=0$ for panels (a) and (b), and $V_0=0.5\, e^2/C_\Sigma$ for panels (c) 
and (d).
} 
\label{res-sens-2d}\end{figure}

Figure \ref{res-sens-2d}(a) shows the RF-SET response $dX/dq_0$  on the plane
of operating point parameters $q_0$ and $V_{in}$ for the case $Q=50$,
$R_\Sigma /R_0=2000$ (i.e.\ $R_\Sigma =100\,\mbox{k}\Omega$), $V_0=0$, 
$\omega =\omega_0$, and $T=0.01 e^2/C_\Sigma$
(the SET is symmetric, $C_1=C_2$, $R_1=R_2$).
Figure \ref{res-sens-2d}(c) is similar, except $V_0=0.5\, e^2/C_\Sigma$. 
 We do not show 
$dY/dq_0$ because it is practically vanishing. Because of the 
same reason, the sensitivity $\delta q_0/\sqrt{\Delta f}$ shown in
Figs.\ \ref{res-sens-2d}(b) and \ref{res-sens-2d}(d) 
is calculated using only quadrature $X$.

Both the response and sensitivity are obviously poor when the amplitude 
of oscillations at the SET is below the Coulomb blockade threshold $V_t$ 
[this happens at $V_{in} < V_t/2Q$ where $V_t=(e/C_\Sigma)(1-2 q_0/e)$, 
and corresponds to the triangles in the lower left corners of Figs.\ 
\ref{res-sens-2d}(a)--(d)], 
and they are also poor
when $V_{in}$ is much larger than this condition (notice that better  
response corresponds to larger $dX/dq_0$, while better sensitivity
corresponds to smaller $\delta q_0$). Even though the regions of
relatively good response and sensitivity are similar, the maximum 
response and optimum sensitivity are achieved at quite different
points in the $V_{in}-q_0$ plane. In particular, the amplitude $V_{in}$ of 
the incoming wave is significantly larger for maximum response, than
for best sensitivity. 

In the present-day experiments, maximization of response 
[Figs.\ \ref{res-sens-2d}(a) and \ref{res-sens-2d}(c)] is still 
of the major importance, because the noise from the next amplifying
stage is still significant [if a large constant noise had been added to $S_X$
in Eq.\ (\ref{dq0X}), then the sensitivity would be mainly determined by the 
denominator $dX/dq_0$]. However, if the amplifier noise is small compared
to the contribution from the SET shot noise, then the best operation 
point should optimize the shot-noise-limited sensitivity
[Figs.\ \ref{res-sens-2d}(b) and \ref{res-sens-2d}(d)]. 
In the case of comparable contributions from two noises, we have a trade-off 
between two regimes. In this paper we will concentrate on the analysis 
of the maximum response (MR) mode and optimized sensitivity (OS) mode,
keeping in mind that experimentally optimal regime is somewhere in between,
depending on the amplifier noise.

\begin{figure}[t]
\centerline{ 
\epsfxsize=3.0in
\epsfbox{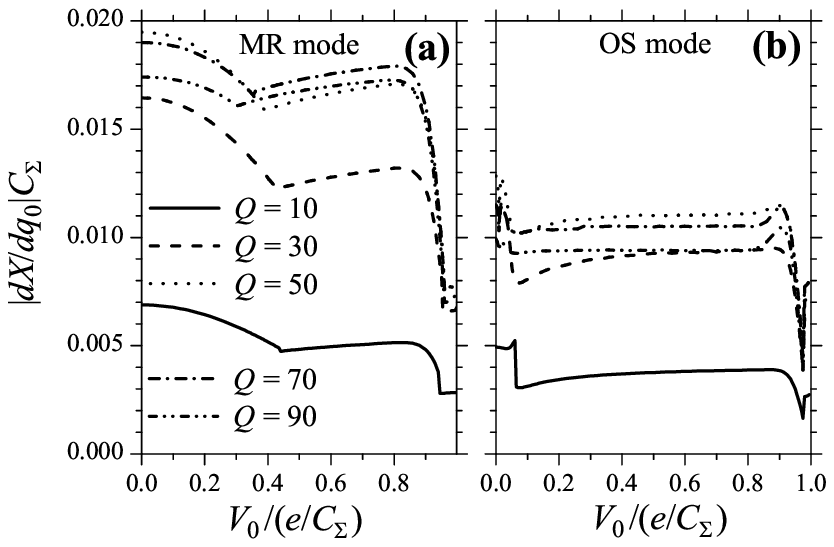} 
}
\centerline{ 
\epsfxsize=3.0in
\epsfbox{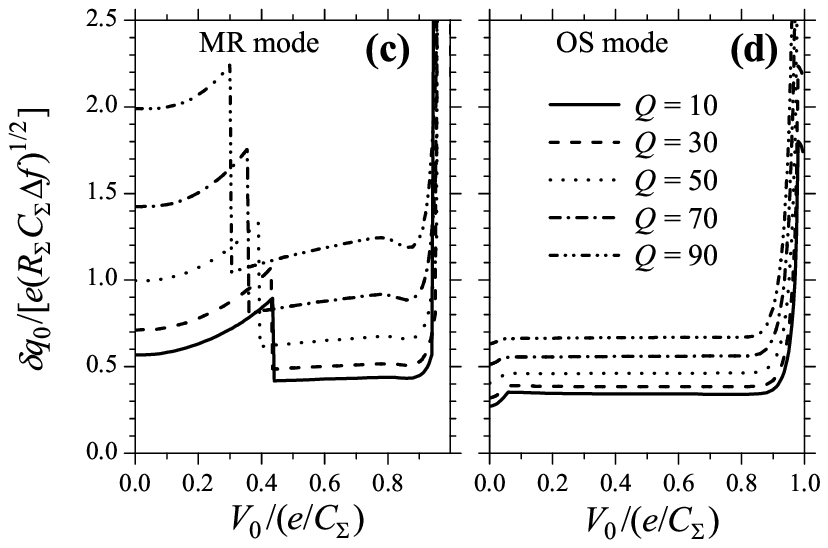} 
}
\vspace{0.3cm}
\caption{Dependence of (a)--(b) the RF-SET response $dX/dq_0$ and
(c)--(d) sensitivity $\delta q_0/\sqrt{\Delta f}$ on the dc bias
voltage $V_0$ for several values of the tank circuit $Q$-factor. 
$V_{in}$ and $q_0$ 
are optimized for either maximum response [MR mode, panels (a) and (c)]
or optimal sensitivity [OS mode, panels (b) and (d)], while other 
parameters are: $T=0.01\, e^2/C_\Sigma$,
$R_\Sigma /R_0=2000$, and $\omega =\omega_0$. 
}
\label{V0dep}\end{figure}

        Figure \ref{V0dep} shows dependence of the response $dX/dq_0$
and sensitivity $\delta q_0/\sqrt{\Delta f}$ on the dc bias voltage $V_0$ 
in the MR and OS modes (optimizations are over $V_{in}$ and $q_0$). 
Several curves on each plot are for different
$Q$-factors: $Q=10$, 30, 50, 70, and 90, while other parameters are
similar to the parameters of Fig.\ \ref{res-sens-2d}.
One can see that the best
response in both MR and OS regimes as well as the best sensitivity
in OS mode are achieved at $V_0=0$. This is because
both positive and negative branches of the symmetric SET $I-V$ curve 
(see Fig.\ \ref{I-V} in Appendix) contribute equally at $V_0=0$,
and therefore the signal is maximal.

 In the MR mode at $V_{0}=0$ the optimum background charge
[see Fig.\ \ref{res-sens-2d}(a)] is about $q_0\approx 0.15\, e$ 
(so the Coulomb blockade threshold $V_t$ is about $0.7\, e/C_\Sigma$) 
while the optimum amplitude $A_b$ of the SET bias $V_b$ oscillations 
is about $1.1\, e/C_\Sigma$
(these numbers have only weak dependence on $Q$, while the optimum $V_{in}$ 
obviously depends on $Q$ quite significantly). 
When $V_{0}$ starts to increase, it becomes advantageous to increase $q_0$ 
(so $V_t$  
decreases) while $A_b$ stays approximately constant, so that both positive
and negative branches of the SET $I-V$ still contribute to the response. 
However, since 
these branches cannot both contribute in the optimal way, the response
decreases with $V_0$ [see Fig.\ \ref{V0dep}(a)]. 
For large enough $V_0$ it becomes preferable to use
only one (positive) branch
[this regime corresponds to the lower maximum in Fig.\ \ref{res-sens-2d}(c),
while the upper maximum corresponds to the two-branch regime]; 
then the optimal wave amplitude drops to
$A_b\simeq 0.2\, e/C_\Sigma$ and the optimal $q_0$ corresponds to $V_t$ 
slightly above $V_0$ (by about $0.03\, e/C_\Sigma$). 
   This change causes the kinks on the response curves in 
Fig.\ \ref{V0dep}(a) and jumps down  on the sensitivity curves in Fig.\
\ref{V0dep}(c). Notice that in the MR mode the sensitivity 
at $V_0\simeq 0.5 \, e/C_\Sigma$ is better than at $V_0=0$. 

        In the OS regime the optimal amplitude is significantly less 
than in the MR regime. It depends mainly on temperature, and for parameters 
of Fig.\ \ref{V0dep} the amplitude $A_b$ at the SET is typically between
0.08 and 0.1 in units of $e/C_\Sigma$. The ``above blockade'' voltage
$V_0+A_b-V_t$ is few times smaller than $A_b$ and is comparable to the 
temperature.
The best sensitivity is achieved at $V_0=0$ when both branches participate,
while at large enough $V_0$ when it becomes preferable to use only one branch,
both sensitivity and response practically do not depend on $V_0$
[see Figs.\ \ref{V0dep}(b) and \ref{V0dep}(d)].

\subsubsection*{Dependence on $Q$-factor}

\begin{figure}[t] 
\centerline{ 
\epsfxsize=2.7in
\epsfbox{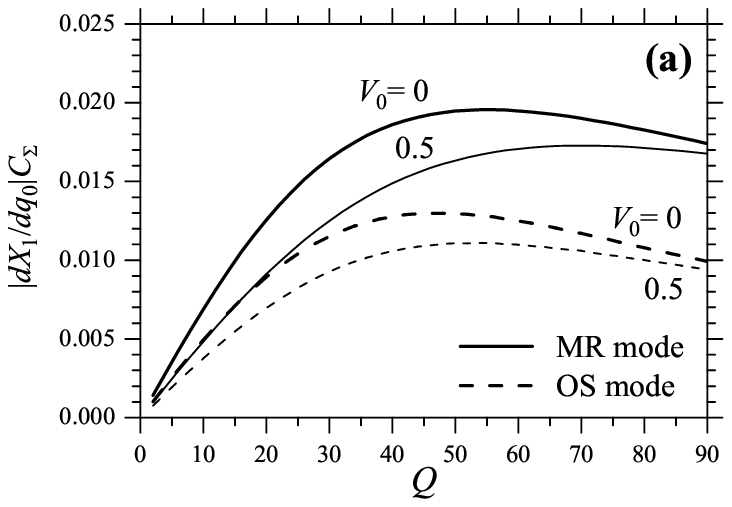} 
}
\vspace{0.1cm}
\centerline{ 
\epsfxsize=2.7in
\epsfbox{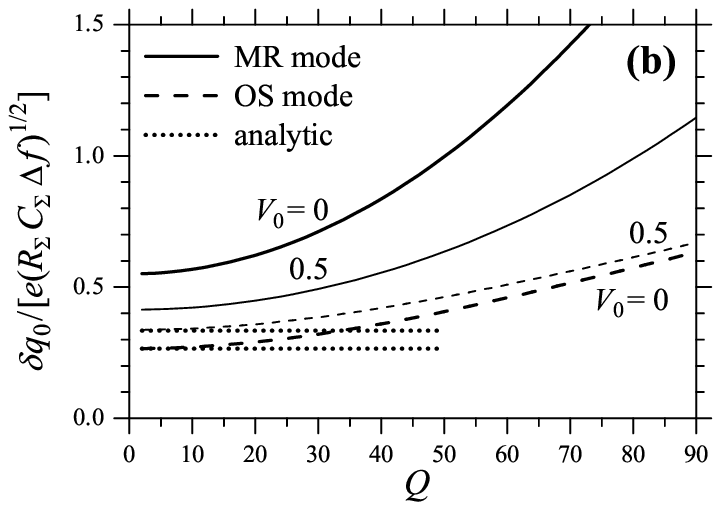} 
}
\vspace{0.2cm}
\caption{(a) RF-SET response and (b) sensitivity 
as functions of the $Q$-factor in the maximum response and optimal
sensitivity modes. $T=0.01\, e^2/C_\Sigma$, $R_\Sigma /R_0=2000$,
and $\omega =\omega_0$. Notice monotonic worsening of the sensitivity
with $Q$. The horizontal dotted lines in (b) show the low-$Q$ low-$T$ 
results \protect\cite{Kor-RF} corresponding to 
 Eqs.\ (\ref{dq-an}) and (\ref{dq-an-2}).  
}
\label{Qdep}\end{figure}

        The dependence on $Q$-factor is summarized in Fig.\ \ref{Qdep}, 
which shows the 
response and sensitivity in the MR and OS modes for $T=0.01\, e^2/C_\Sigma$,
$R_\Sigma /R_0=2000$, and $\omega =\omega_0$. The results are presented
for $V_0=0$ (thick lines), which provides the best MR and OS response 
and best OS sensitivity, and also for 
$V_0=0.5\, e/C_\Sigma$ (thin lines), which is a typical value for the 
case when only
one branch of the SET $I-V$ is involved.
 One can see that the response $dX/dq_0$ grows linearly with $Q$ at 
small $Q$ [see Eq.\ (\ref{resp-lin})] and 
reaches the maximum 
at $Q$ around 50 (this number is somewhat different for different regimes;
for example it is almost 70 for the thin solid line), 
which is close to the crude theoretical estimate $\sqrt{R_\Sigma /R_0}
\simeq 45$ for the impedance matching. 
However, unlike in the linear model, this maximum does not 
correspond to the exact impedance matching. For example, 
the impedance matching (minimum of reflection) occurs at $Q\simeq 100$ 
for the upper curve in Fig.\ \ref{Qdep}(a) and at $Q\simeq 80$ for 
the curve second from the top, while for two lower curves (OS mode)
it does not occur at all in a reasonable range of $Q$.

In contrast to the response behavior, {\it the RF-SET sensitivity 
monotonically worsens with increase of} $Q$. 
Qualitatively, this happens because the noise $S_X$ has $Q^2$ scaling 
[see Eq.\ (\ref{SX1})], while the response $dX/dq_0$ has slower than
$Q$ dependence [see Eq.\ (\ref{resp-lin})]. This simple analysis 
predicts the parabolic dependence $\delta q_0 \propto 1+Q^2/(R_d/R_0)$, 
which crudely fits the curves in Fig.\ \ref{Qdep}(b) using 
$R_d/R_0$ about 1.5-3 times larger than $R_\Sigma /R_0$, though the 
curves in Fig.\ \ref{Qdep}(b) actually have slower than parabolic
dependences at large $Q$.

        Comparing the MR and OS modes at $Q=30$ (just some
typical number) and $V_0=0$, we see that the MR regime provides about 40\%
larger response, while the OS regime provides about twice better sensitivity.
Even though these numbers depend significantly  on the temperature
and also depend on $Q$ and $R_\Sigma /R_0$, they show that the results for 
the MR and OS regimes are not too much different (not by an order of 
magnitude). 

Comparison of the cases $V_0=0$ and $V_0=0.5\, e/C_\Sigma$ shows that for 
the MR and OS response as well as for the OS sensitivity there is no much
difference between 
these two cases, and the relative difference decreases with $Q$. 
In contrast, the difference between MR sensitivity at $V_0=0$ and
at $V_0=0.5\, e/C_\Sigma$ grows with $Q$ and can become significant.

The low-$Q$ limit of the OS sensitivity for $V_0=0$ is well described by 
the formula \cite{Kor-RF} (which also assumes $T\ll e^2/C_\Sigma$) 
        \begin{equation}
        \delta q_0/\sqrt{\Delta f} \simeq 2.65\, e(R_\Sigma C_\Sigma)^{1/2} 
                (TC_\Sigma /e^2)^{1/2} 
        \label{dq-an}\end{equation}
[we do not use a shorter formula $2.65\, C_\Sigma (R_\Sigma T)^{1/2}$ to
emphasize natural normalizations], and similar limit for $V_0=0.5 \, 
e/C_\Sigma$ (one-branch case) is close to \cite{Kor-RF}
        \begin{equation}
        \delta q_0/\sqrt{\Delta f} \simeq 3.34\, e(R_\Sigma C_\Sigma)^{1/2} 
                (TC_\Sigma /e^2)^{1/2}  
        \label{dq-an-2}\end{equation}
[see Fig.\ \ref{Qdep}(b)]. 
However, the theory of Ref.\ \cite{Kor-RF} which assumes 
$Q\ll \sqrt{R_\Sigma /R_0}$, is not able to describe significant change of
the sensitivity with $Q$ in Fig.\ \ref{Qdep}(b). 
One can see that this dependence is even more significant in the MR mode.

\begin{figure}[t]
\centerline{ 
\epsfxsize=3.0in
\epsfbox{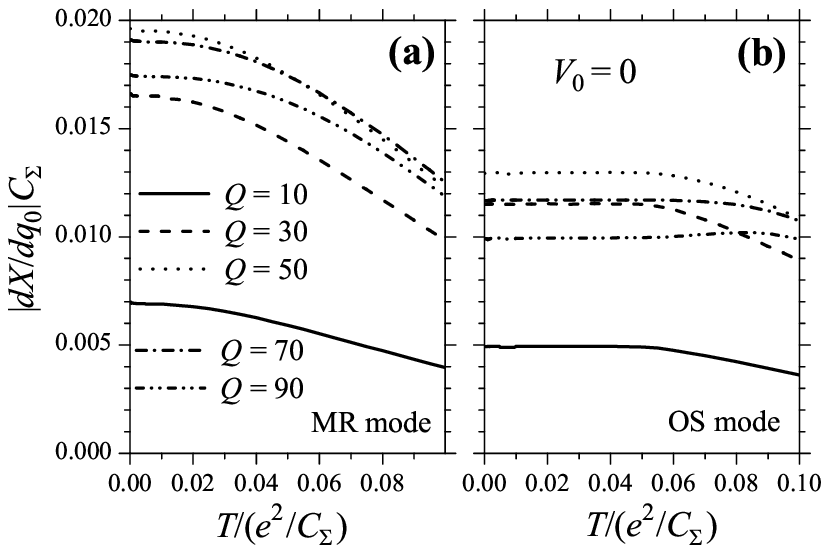} 
}
\centerline{ 
\epsfxsize=3.0in
\epsfbox{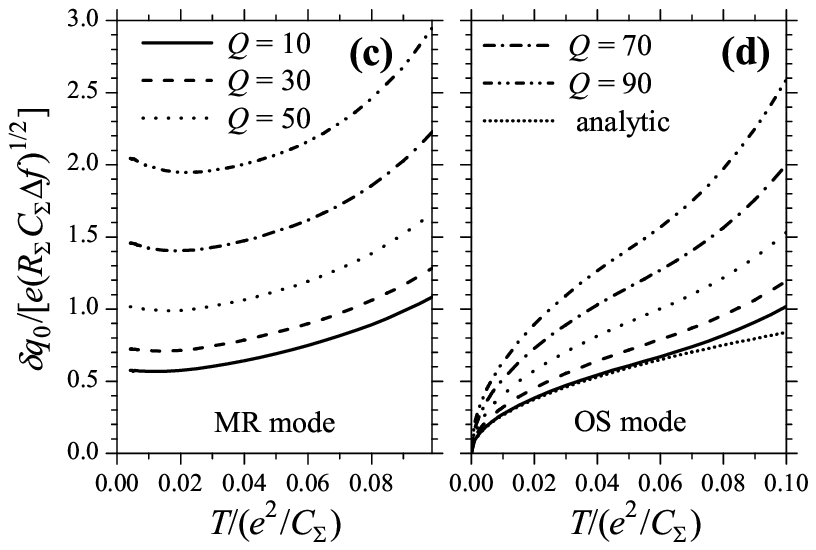} 
}
\vspace{0.3cm}
\caption{Dependence of (a)--(b) RF-SET response and (c)--(d) sensitivity
on temperature $T$ in the MR and OS regimes for several $Q$-factors. 
$V_0=0$, $R_\Sigma /R_0=2000$, and $\omega =\omega_0$. Lowest dotted line 
in (d) is Eq.\ (\ref{dq-an}).
}
\label{Tdep}\end{figure}

\subsubsection*{Dependence on temperature and SET resistance} 

        The numerical results on temperature dependence are shown in
Fig.\ \ref{Tdep} for $R_\Sigma /R_0=2000$ and $\omega =\omega_0$. 
Similar to Fig.\ \ref{V0dep} we show on four panels
the response $dX/dq_0$ and corresponding sensitivity in the MR and OS
modes (the optimal dc bias value $V_0=0$ is used). 
It is important to notice that in the MR mode both response 
and sensitivity almost do not depend on temperature at 
$T < 0.03 \, e^2/C_\Sigma$, and the RF-SET performance is still OK 
at temperatures $\sim 0.1\, e^2/C_\Sigma$ (response and sensitivity
change less than twice compared to zero-temperature case). 
In the OS regime the response also has a very weak temperature dependence 
at $T<0.05 e^2/C_\Sigma$; however, the sensitivity is strongly
temperature-dependent. The low-$Q$ OS sensitivity can be accurately 
described by Eq.\ (\ref{dq-an}) up to temperatures $\sim 0.05\, e^2/C_\Sigma$,
[see the lowest dotted line in Fig.\ \ref{Tdep}(d)],
and the curves for large $Q$-factors also follow the 
scaling $\delta q_0/\sqrt{\Delta f} \propto T^{1/2}$ at temperatures
$T < 0.05\, e^2/C_\Sigma$. In the orthodox theory this $T^{1/2}$ dependence 
is valid even at very small temperatures leading to infinitely good 
sensitivity ($\delta q_0/\sqrt{\Delta f}\rightarrow 0$); 
however, in reality the neglected contribution of cotunneling processes 
becomes significant in the OS regime at small $T$ that changes 
the formalism.\cite{Kor-92,Devoret}
    Comparing the results for MR and OS modes, we notice that while they 
are significantly different at low temperatures, the difference 
decreases with temperature, so that at $T\sim 0.1\, e^2/C_\Sigma$
the MR and OS results are already quite similar.

\begin{figure}[t]
\centerline{ 
\epsfxsize=3.0in
\epsfbox{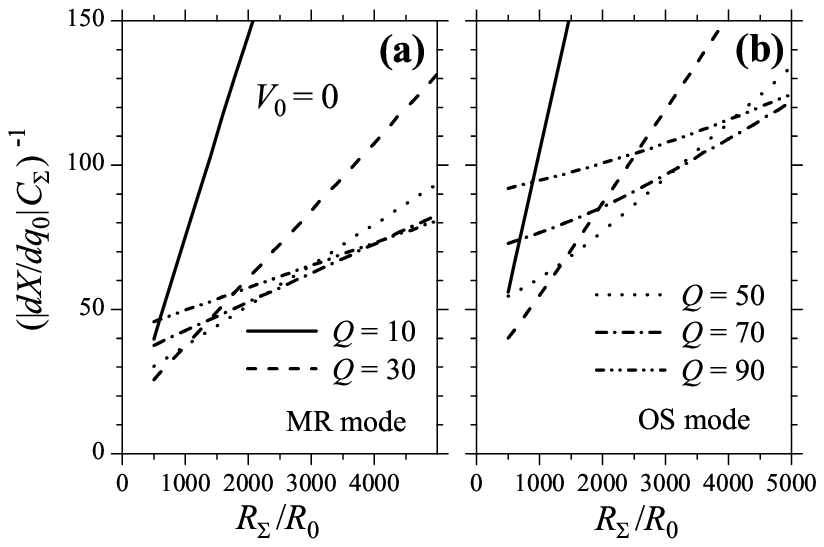} 
}
\centerline{ 
\epsfxsize=3.0in
\epsfbox{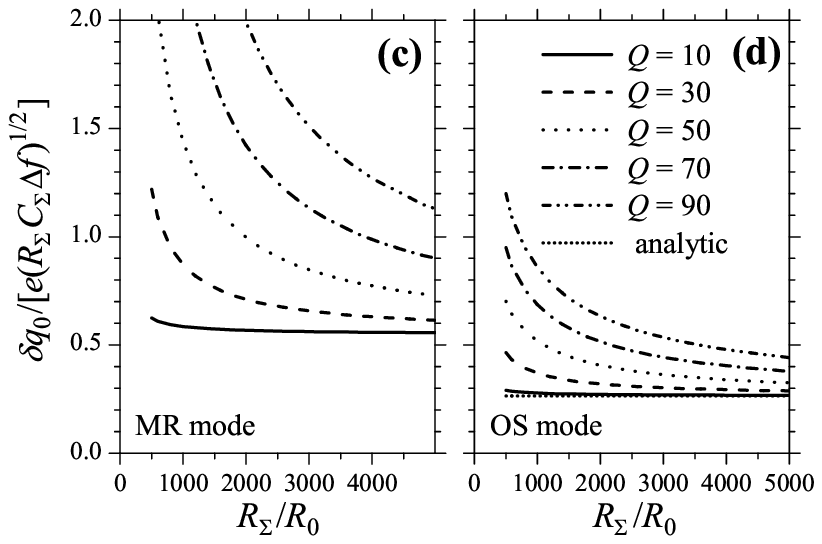} 
}
\vspace{0.3cm}
\caption{(a)--(b) Inverse response and (c)--(d) sensitivity in the 
MR and OS regimes as functions of the SET resistance $R_\Sigma$ for several
$Q$-factors at $T=0.01$, $V_0=0$, and $\omega =\omega_0$. 
}
\label{Rdep}\end{figure}

    Figure \ref{Rdep} shows the response and sensitivity dependence 
on the total SET junction resistance $R_\Sigma$ for $T=0.01\, e^2/C_\Sigma$,
$V_0=0$, 
and $\omega =\omega_0$ (we show results only for $R_\Sigma /R_0>500$,
because at $R_\Sigma < 25 \, \mbox{k}\Omega$ the theory is too inaccurate
due to neglected cotunneling processes). 
 In the case \cite{Kor-RF} $Q\ll \sqrt{R_\Sigma /R_0}$ 
the response $dX/dq_0$ scales as $R_\Sigma^{-1}$ [see Eq.\ (\ref{X1})]
and the sensitivity $\delta q_0/\sqrt{\Delta f}$ scales as $R_\Sigma^{1/2}$
[see Eq.\ \ref{dq-an}]. 
Correspondingly, the solid lines ($Q=10$) in Figs.\ \ref{Rdep}(a) and (b)
are practically straight lines passing through the origin, and
in Figs.\ \ref{Rdep}(c) and (d) the solid lines are
practically horizontal [the level determined by Eq.\ (\ref{dq-an})
is shown in Fig.\ \ref{Rdep}(d) by the lowest dotted line, which fits 
well the solid line]. 
  The dependence on $R_\Sigma$ becomes nontrivial
when $R_\Sigma /R_0$ is comparable to $Q^2$. It is interesting to notice
that the inverse response remains a practically linear function of 
$R_\Sigma$ even for large $Q$, as seen in Figs. \ref{Rdep}(a) and (b)
(the slope of the lines decreases with $Q$ while the offset increases). 
In particular, this means that in the orthodox model the RF-SET response 
continues to increase with  the SET resistance decrease 
even when the matching condition is overshot.
Decrease of $R_\Sigma$ close or beyond the matching condition, worsens
the sensitivity in comparison with scaling 
$\delta q_0/\sqrt{\Delta f} \propto R_\Sigma^{1/2}$ [see Figs.\
\ref{Rdep}(c) and (d)]. In the MR mode the sensitivity worsens with decrease
of $R_\Sigma$ even in absolute units ($e/\sqrt{\mbox{Hz}}$) when the matching 
condition is sufficiently overshot, while in the OS mode the sensitivity
still improves with decrease of $R_\Sigma$ in absolute units, in spite
of worsening compared to $R_\Sigma^{1/2}$ scaling.

\subsubsection*{Effect of asymmetric biasing ($C_1\neq C_2$)} 

\begin{figure}[t]
\centerline{ 
\epsfxsize=3.0in
\epsfbox{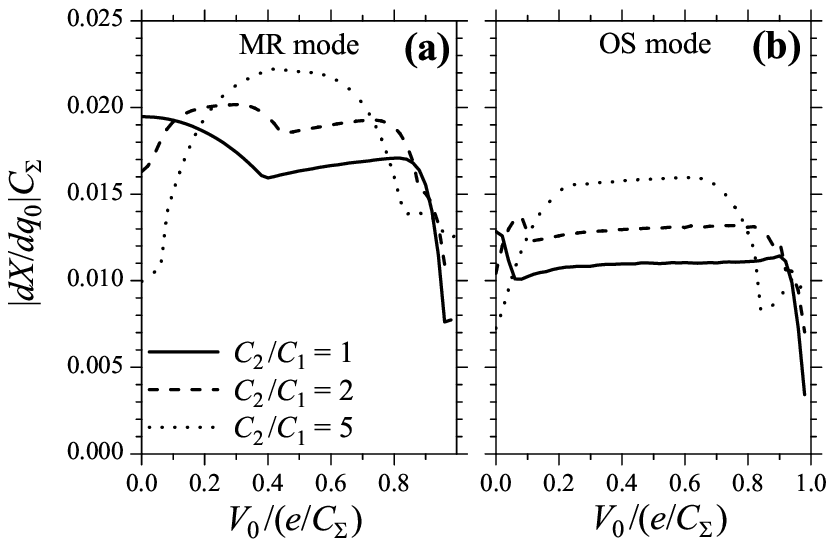} 
}
\centerline{ 
\epsfxsize=3.0in
\epsfbox{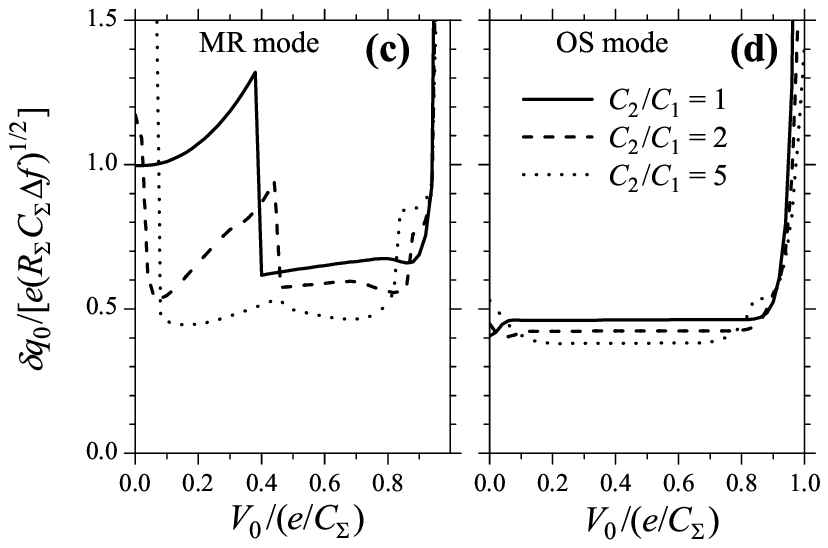} 
}
\vspace{0.3cm}
\caption{(a)--(b) Dependence of the RF-SET response and (c)--(d) sensitivity 
on the dc bias voltage $V_0$ for several values of the effective asymetry 
of the SET capacitances. 
$T=0.01 e^2/C_\Sigma$, $Q=50$, $R_\Sigma /R_0=2000$, and $\omega =\omega_0$.
Notice that the SET asymmetry (``asymmetric biasing'') slightly improves 
the RF-SET performance. 
}
\label{asymbias}\end{figure}

        Now let us discuss the effect of asymmetric effective capacitances
of the SET junctions, $C_1\neq C_2$. Even when the physical junction 
capacitances are equal, $C_{1j}=C_{2j}$, the effective capacitances 
$C_1$ and $C_2$ can be significantly different when the gate capacitance
$C_g$ is comparable to junction capacitances. 
 In our model shown in Fig.\ \ref{fig1} this happens due to asymmetry 
of signal source capacitances, $C_{S1}\neq C_{S2}$ [see Eqs.\ 
(\ref{C1}) and (\ref{C2})]; in RF-SET experiments this effect is 
called asymmetric rf biasing of the SET.\cite{Aassime} 
For the conventional SET setup, the biasing asymmetry is not important at 
all, because of the formal SET equivalence under transformation 
\cite{Kor-ff}
        \begin{equation}
C_1\rightarrow C_1+\Delta C, \,\,
C_2\rightarrow C_2-\Delta C, \,\,
q_0\rightarrow q_0 - V_b \Delta C 
        \end{equation} 
for arbitrary $\Delta C$, that means that the asymmetric biasing
(different $C_1$ and $C_2$) can be simply corrected by the
background charge shift. However, for the RF-SET there is no simple 
equivalence because the bias voltage $V_b$ changes in time. (Effective 
capacitance asymmetry can still be controlled by addition of the extra 
rf signal to the SET gate.) Therefore it is interesting to find if the
asymmetric biasing ($C_1\neq C_2$) is better or worse than the symmetric 
case ($C_1=C_2$).

   Figure \ref{asymbias} shows the $V_0$ dependence of the RF-SET response
and sensitivity in the MR and OS modes for several ratios $C_1/C_2$
(with fixed total capacitance $C_1+C_2$). One
can see that in the asymmetric cases the best MR response and OS sensitivity
are achieved at nonzero $V_0$ and are typically $\it better$ than the
corresponding values for the symmetric case. Therefore, the {\it asymmetric
biasing is actually preferable} (that eases the concern about this issue 
discussed in Ref.\ \cite{Aassime}); however, the advantage is rather minor.
In particular, the OS sensitivity in the asymmetric case is still limited
by Eq.\ (\ref{dq-an-2}); therefore the possible improvement is less than 
the sensitivity decline due to high $Q$-factor [see Fig.\ \ref{Qdep}(b)],
and even less because of the difference between Eqs.\ (\ref{dq-an}) and 
(\ref{dq-an-2}).

\subsubsection*{Carrier frequency detuning from the resonance} 

        So far we have considered the resonant case $\omega =\omega_0$.
[Actually, the exact resonance is at the frequency 
$\omega =\omega_0 (1-1/2Q_{SET}^2)$ which is very close to $\omega_0$.] 
In this case the quadrature component $X$ is much larger than $Y$ 
(at $Q\gg 1$) and therefore the RF-SET response and sensitivity in respect 
to monitoring $X$ quadrature practically coincide with that
for monitoring the reflected wave amplitude $A$ (we denote with $A$ the 
amplitude $A_1$ of the first harmonic) or monitoring the optimized 
phase-shifted combination $X^* =X\cos \varphi +Y \sin\varphi$. 
The detuning of $\omega$ from the resonant frequency $\omega_0$ leads 
to a significant magnitude of $Y$ quadrature, and so to different
results for different ways of reflected wave monitoring.

For a small frequency detuning $\Delta \omega =\omega -\omega_0$ 
the simple linear analysis using Eq.\ (\ref{resp-lin}) predicts 
the non-monotonic shape of the $X$-response frequency dependence 
with zeros at $\omega /\omega_0 =1\pm 1/2Q_L$ and with the full width 
at half height (FWHH) equal to $0.53 \omega_0/Q_L$; while 
for the amplitude monitoring it predicts FWHH of $\sqrt{3} \omega_0/Q_L$.
[Eq.\ (\ref{dadr}) which assumes both $R_d$ and $V_{in}$ being 
frequency-independent, gives the response FWHH equal to 
$(\sqrt{5}-2)^{1/2}\omega_0/ Q_L$ and $\omega_0/Q_L$ for $X$ and $A$
monitoring, respectively]. 
In reality the situation is more complex because even neglecting
the nonlinear contributions, the effective SET resistance 
$R_d$ (and therefore $Q_L$) depends on $V_b(t)$ amplitude, 
which depends itself on $\omega$, 
$R_d$, and $V_{in}$. Moreover, the optimized point in the $V_{in}-q_0$
plane is also frequency-dependent.

\begin{figure}[t]
\centerline{ 
\epsfxsize=3.0in
\epsfbox{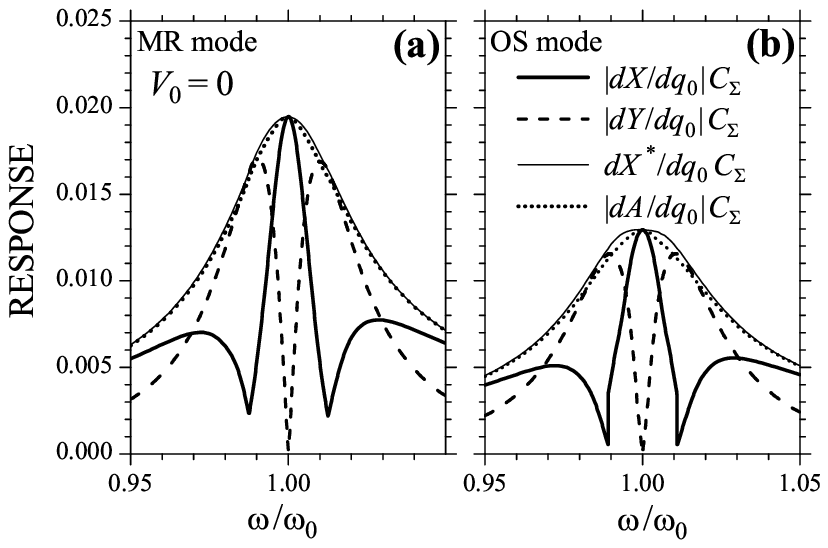} 
}
\centerline{ 
\epsfxsize=3.0in
\epsfbox{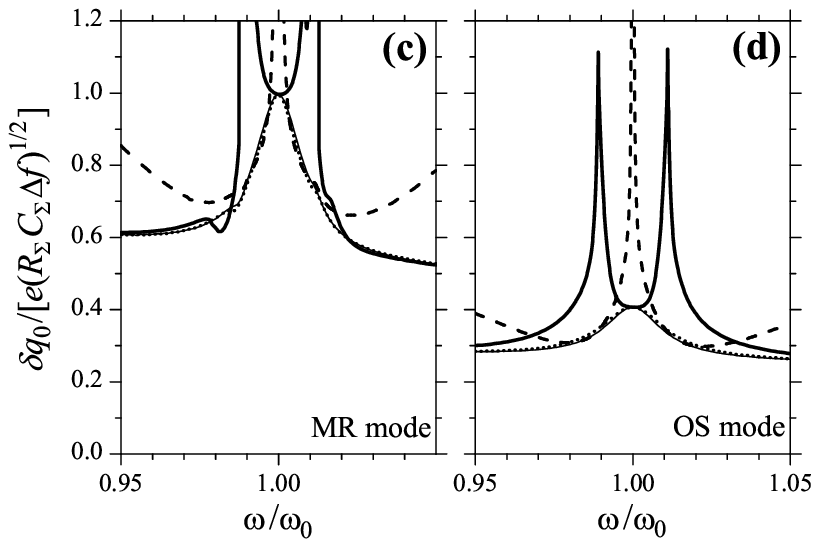} 
}
\vspace{0.3cm}
\caption{Frequency dependence of the (a)--(b) RF-SET response and
(c)--(d) sensitivity in the MR and OS modes. Each panel shows four 
curves corresponding to monitoring of $X$ quadrature (thick solid line), 
$Y$ quadrature (dashed), their optimized phase-shifted combination
$X^*$ (thin solid), and the first harmonic amplitude $A$ (dotted). 
$Q=50$, $T=0.01 e^2/C_\Sigma$, $R_\Sigma /R_0 =2000$, $V_0=0$. 
}
\label{freq-dep}\end{figure}

        Figure \ref{freq-dep} shows the numerical results for the 
frequency dependence (in the vicinity of the resonance) of the 
optimized response and sensitivity. Each panel shows four curves, which 
correspond to monitoring the components $X$ (thick solid line), $Y$ (dashed),
optimized phase-shifted combination $X^*$ (thin solid), 
and amplitude $A$ (dotted line). 
At each frequency we perform optimizations of the response (MR mode)
and sensitivity (OS mode) over $V_{in}-q_0$ plane for all four 
monitored magnitudes, so the optimizations are different for different 
curves.
The $Q$-factor is equal to $50$, which is close to the 
estimate of the impedance-matching value (see also Fig.\ \ref{Qdep});
however, instead of the naively expected value $Q_L \simeq Q/2= 25$ 
for the loaded $Q$-factor, it is $Q_L = 40$ in the MR mode since 
$Q_{SET}=199$ (these values are calculated at $\omega =\omega_0$ using
effective SET resistance). The shape of the $X$-response dependence 
in panel (a) is close to the prediction of Eq.\ (\ref{resp-lin}) 
using $Q_L=40$, though the minima do not reach zero and the curve beyond
the minima is shifted up. 
The MR $Y$-response is practically zero at $\omega =\omega_0$ (since $Y$
vanishes); however, it becomes comparable to the resonant $X$-response 
at some frequency detuning; the overall shape is close to the prediction 
of Eq.\ (\ref{resp-lin}), but the maxima are about 15\% higher. 
The MR $A$-response curve is about 50\% wider than the prediction of the 
linear theory (using $Q_L=40$) 
and is very close to the curve for monitoring optimal
phase-shifted quadrature $X^*$. As expected, the $A$- and $X^*$-responses 
are better than $X$- and $Y$-responses at finite detuning (for both
MR and OS modes). 

        The sensitivity (both OS and MR) for $X^*$ or  
$A$ monitoring at finite detuning is also better then for 
$X$ or $Y$ monitoring. An interesting 
observation is that while the RF-SET response decreases with detuning, 
the sensitivity slightly improves with detuning for $X^*$ and $A$ monitoring.
This effect is similar to the sensitivity improvement
with the decrease of $Q$-factor (see Fig.\ \ref{Qdep}).

\subsubsection*{Monitoring of resonant overtone} 

        In experiments the incoming rf wave is usually tuned close
to the resonance with the tank circuit; in this case the contribution of 
overtones in the reflected rf wave is small in spite
of significant SET $I-V$ nonlinearity (the SET nonlinearity has been 
recently used \cite{Knobel} for rf mixing). 
    However, if the $n$th overtone is in resonance, 
$\omega \approx \omega_0/n$, then the reflected wave may have a significant 
contribution from this overtone, and the RF-SET operation can be based 
on monitoring this overtone. \cite{Turin-Kor} The use of different 
frequencies for the incident and reflected waves may be advantageous 
for some applications. 
Also, it may be useful to have the absence of the monitored reflected wave 
when the SET is off (no current), while in the conventional regime this case 
corresponds to the largest reflected power. 
One more possible advantage is somewhat easier control of the amplitude 
of the SET bias voltage 
oscillations, since now it is more directly related to the incident amplitude
$V_{in}$, while in the usual regime the relation depends on $Q_L$
[see Eq.\ (\ref{Vbsimple})] which varies with operating point.
(The disadvantage is that the incident amplitude $V_{in}$ should be much 
larger than in usual regime that may lead to heating problems.)

\begin{figure}[t]
\centerline{ 
\epsfxsize=2.7in
\epsfbox{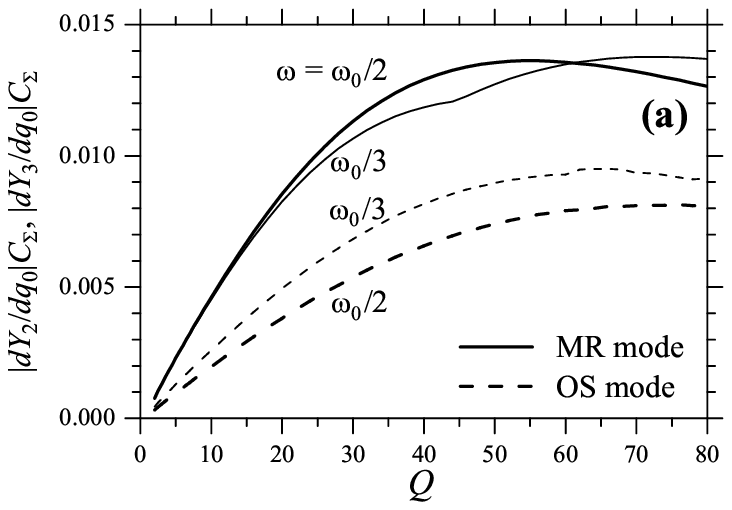} 
}
\centerline{ 
\epsfxsize=2.7in
\epsfbox{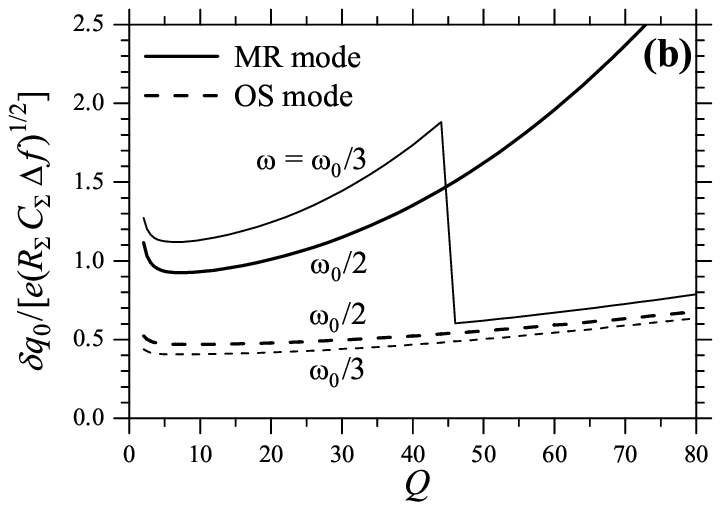} 
}
\vspace{0.3cm}
\caption{(a) RF-SET response and (b) sensitivity in the regimes when the 
second or third overtone of the incident rf wave is in resonance with 
the tank circuit. 
$T=0.01\, e^2/C_\Sigma$, $R_\Sigma /R_0=2000$;
$V_0=0.5e/C_\Sigma$ for $\omega =
\omega_0/2$ and $V_0=0$ for $\omega =\omega_0/3$. 
}
\label{overtones}\end{figure}

        Figure \ref{overtones} shows the RF-SET response and sensitivity
for $\omega =\omega_0/2$ and $\omega =\omega_0/3$, in respect 
to monitoring $Y_2$ and $Y_3$, 
correspondingly (the $X$-quadratures are small). We use $V_0=0$
in the case $\omega =\omega_0/3$ and $V_0=0.5 e/C_\Sigma$  in the case 
$\omega =\omega_0/2$ [for $V_0=0$ there is no second overtone because
of the $I-V$ curve symmetry even at nonzero $q_0$ -- see Fig.\ \ref{I-V}(a) 
and Eq.\ (\ref{anbn})]. 
As seen from Fig.\ \ref{overtones}, the MR responses and OS sensitivities 
in the two cases 
(second and third resonant overtone) are quite similar, that is related
to a strong nonlinearity of the SET $I-V$ curve. 
Moreover, comparing Figs.\ \ref{Qdep} and \ref{overtones} (the parameters 
are the same) we see that 
the RF-SET performance in the regime of a resonant 
overtone is comparable to the performance in the conventional regime $\omega 
= \omega_0$ (the MR response and OS sensitivity are worse by only about  
1.5 times). Combined with the advantages discussed above, this makes 
the regime of resonant overtone potentially useful in experiments.

\section{Conclusion}

        In this paper we have developed the formalism for the
calculation of the response and shot-noise-limited sensitivity
of the normal-metal RF-SET, extending the theory of Ref.\ \cite{Kor-RF}
to the case of arbitrary large $Q$-factor of the tank circuit and 
arbitrary frequency of the incident rf wave. The formalism has been
used to analyze numerically the dependence of the RF-SET response
and sensitivity on the operation parameters.

We have mainly studied two operation modes, optimized over the rf wave
amplitude and the SET background charge, corresponding to either 
maximum response (MR mode) or optimum sensitivity (OS mode). The rf 
amplitude for the optimum sensitivity is typically significantly smaller 
than for maximum response. The MR mode is the best experimental regime
when the preamplifier noise is more significant than the effect of the SET 
noise, while the OS mode is the best when the preamplifier noise is
negligible. 
        Analyzing the performance dependence on the SET dc bias voltage, 
we have found that the best response and sensitivity are achieved
at zero dc bias, though finite biasing does not change the RF-SET 
performance much as long as it is within the Coulomb blockade range. 

        We have found that the dependence of the RF-SET response 
on the unloaded $Q$-factor of the tank circuit has a maximum at $Q$ 
comparable to the simple impedance matching estimate $\sqrt {R_\Sigma /R_0}$.
In contrast, the RF-SET sensitivity monotonically worsens with increase
of $Q$; the dependence is approximately parabolic and the sensitivity 
change can reach few times (compared to the low-$Q$ case) if $Q$ is chosen 
too large. This means that to improve the sensitivity in an experiment, 
it is better to ``undershoot'' the $Q$-factor compared to the impedance
matching case, than to ``overshoot'' it. 

        Studying the temperature dependence we have found that the RF-SET
response saturates approximately at temperatures $T < 0.03\, e^2/C_\Sigma$,
that translates into 200 mK for a typical capacitance value 
$C_\Sigma = 300$ aF. 
The optimized sensitivity continues to improve as $T^{1/2}$ at lower 
temperatures until it reaches the quantum limitation due to cotunneling 
\cite{Kor-92,Devoret} (not studied here). The ``orthodox'' RF-SET 
performance improves with decrease of the SET resistance; however,
for high $Q$-factors the dependence is significantly slower than
the low-$Q$ scaling $R_\Sigma^{-1}$ for the response and $R_{\Sigma}^{1/2}$
for the sensitivity.

        We have analyzed the effect of the asymmetric rf biasing of the SET
which leads to unequal effective capacitances of the SET and found
that such asymmetry does not worsen the RF-SET performance (even slightly
improves it). This answers the concern about asymmetric rf biasing 
raised in Ref.\ \cite{Aassime}.  

        We have analyzed the effect of the carrier frequency detuning 
from the resonance and found that the decrease of the RF-SET response 
with detuning can crudely be described by the simple linear theory 
(though difference in linewidth can reach 50\%); however, the estimate
of the  loaded $Q$-factor determining the linewidth is not simple 
since it significantly depends of the operation point. Even with the
frequency detuning, the RF-SET performance for monitoring the reflected 
wave amplitude (by rectification) is found to be very similar to 
the case of optimal homodyne detection (one-channel mixing with the optimal
phase). It is important that the mixing phase optimizing the sensitivity
can be different from the phase optimizing the response. 
Unexpectedly, in contrast to the response decrease with detuning, 
the optimized sensitivity slightly improves with detuning. 

        We have also analyzed the operation regime for which an  
overtone of the incident rf wave is in resonance with the tank circuit, 
and found
that the RF-SET performance in this regime is comparable to the performance
in the conventional regime. Taking into account an advantage of the 
frequency separation between the incident and monitored waves, this 
operation mode may be experimentally useful. The theoretical proposal
of the resonant overtone regime \cite{Turin-Kor} has been recently 
realized experimentally in the group of Keith Schwab. Experimental
RF-SET sensitivities using the second and third resonant overtones 
have been found practically coinciding with the sensitivity in the
conventional regime.
\cite{Schwab-overtone} 

        Now let us compare our theoretical results for the RF-SET 
sensitivity with the experimental value of 
$9\, \mu e/\sqrt{\mbox{Hz}}$ from Ref.\ \cite{Aassime-2} 
for the normal-metal case (the sensitivity of the superconducting 
RF-SET was significantly better: $3.2\, \mu e/\sqrt{\mbox{Hz}}$).
Using the experimental parameters $C_\Sigma =267$ aF, 
$R_\Sigma =43$ k$\Omega$,
assuming the temperature $T=70$ mK, and using Fig.\ \ref{Rdep},
we get the MR sensitivity of $3.2\,\mu e/\sqrt{\mbox{Hz}}$
for $Q=30$ and $5.8\,\mu e/\sqrt{\mbox{Hz}}$ for $Q=50$,
while the OS sensitivities are $1.3\,\mu e/\sqrt{\mbox{Hz}}$
for $Q=30$ and $1.9\,\mu e/\sqrt{\mbox{Hz}}$ for $Q=50$
[notice that all these numbers are significantly higher than the 
low-$Q$ OS estimate $0.9\,\mu e/\sqrt{\mbox{Hz}}$  using
Eq.\ (\ref{dq-an})]. 
Unfortunately, it is not possible to extract the unloaded $Q$-factor value 
from Ref.\ \cite{Aassime-2}; however, it seems to be between 30 and 50.
It is also not known if the RF-SET operation point was closer to the MR
or OS mode (we guess the response was more important). 
Nevertheless, we see that the difference between the theory and experiment
is few times. We guess that the difference is mainly due to the 
preamplifier noise. Some contribution may also come from higher effective 
temperature than we assumed and from a non-optimal 
operation point. Another contribution to the difference may
come from the neglected here effect of cotunneling, which limits the 
sensitivity. However, a rough estimate \cite{Kor-92} of this limit  
$\delta q_0 /\sqrt{\Delta f} \sim \sqrt{\hbar C_\Sigma}$ gives the 
value of $1\,\mu e/\sqrt{\mbox{Hz}}$, so it is unlikely to be 
the major reason for the difference. We hope that the further 
experimental progress will bring the RF-SET sensitivity really close
to the theoretical limit. 

        There are still many theoretical questions about the RF-SET 
performance, not answered in this paper. For example, as seen from the
above estimate, the account of cotunneling and study of the quantum
operation of the RF-SET is starting to be important for experiments. 
The development of the theory for superconducting RF-SET is even more
important since in the majority of experiments with RF-SETs the 
superconducting state is used. There is still no rigorous theory 
of the frequency dependence of the RF-SET sensitivity. It is also
important to consider the backaction from the RF-SET and analyze
if it can in principle be used as a quantum detector with high
quantum efficiency (ideality). 
These problems are going to be the topics of further studies.

\section*{Acknowledgments}

Useful discussions with M. Blencowe, K. Likharev, and K. Schwab 
are gratefully acknowledged. 
The work was partially supported by NSA and ARDA under ARO grant 
DAAD19-01-1-0491 and by the Semiconductor Research Corporation grant
2000-NJ-746. The numerical calculations were partially performed 
on the UCR-IGPP Beowulf computer Lupin.


        \begin{appendix}
\section*{Current and noise calculations for the SET}

        The SET current
$I$ and its low frequency shot noise $S_I$ have been calculated using
the orthodox theory of single-electron tunneling,\cite{Av-Likh} 
assuming that a stationary state of the SET is achieved at any
moment of time (i.e.\ $I/e \ll \omega$). We assume normal metal 
(non-superconducting) case and calculate the tunneling rates 
$\Gamma_{1,2}^\pm(m)$ of electron tunneling to ($+$) or from ($-$)
the island through the first or second junction as\cite{Av-Likh} 
        \begin{eqnarray}
&& \Gamma_j^\pm (m) =\frac{W_j^\pm(m)}{e^2R_j[1-\exp (-W_j^\pm (m)/T)]} \, ,
\\
&& W_j^\pm (m) = \frac{e^2}{C_\Sigma} \left[ \mp (-1)^j 
\frac{V_bC_1C_2}{eC_j}
-\frac{1}{2} \mp m \mp \frac{q_0}{e} \right] , 
        \end{eqnarray}
where $m$ is the number of extra electrons on the SET, $j=1,2$ denotes the 
junction, and $T$ is the 
temperature.

\begin{figure}[t]
\centerline{ 
\epsfxsize=2.5in 
\epsfbox{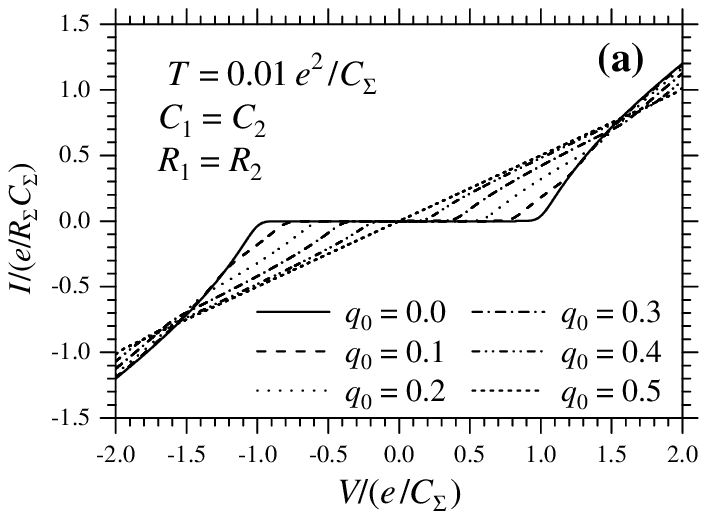} 
}
\centerline{ 
\epsfxsize=2.5in 
\epsfbox{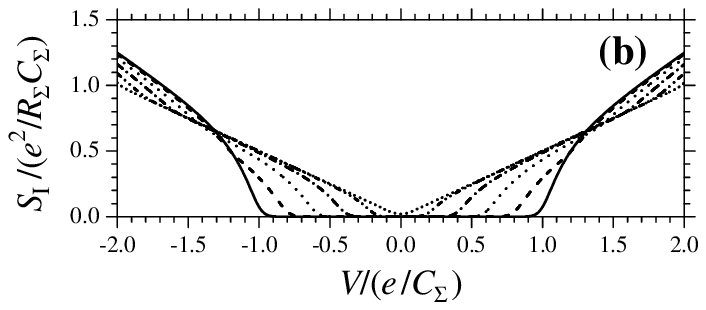} 
}
\vspace{0.3cm}
\caption{(a) $I-V$ curves and (b) voltage dependence of 
low-frequency shot noise $S_I$
 for the symmetric SET at $T=0.01\, e^2/C_\Sigma$
for several values of the background charge $q_0$
from 0 to 0.5 (in units of $e$).
}
\label{I-V}\end{figure}

The current [Fig.\ \ref{I-V}(a)] is calculated as 
        \begin{equation}
I=e\sum_m [\Gamma_1^+(m)-\Gamma_1^-(m)]\sigma_{st}(m),
        \end{equation}
where the stationary probability distribution of the charge states
$\sigma_{st}(m)$
satisfies equations
        \begin{equation} 
\sigma_{st}(m)\sum_{j}\Gamma_j^+(m) = 
\sigma_{st}(m+1)\sum_j\Gamma_j^-(m+1)
        \end{equation}
and $\sum_m \sigma_{st}(m)=1$. 

The low-frequency spectral density $S_I$
of the SET current [Fig.\ \ref{I-V}(b)] can be calculated as \cite{Kor-94}
        \begin{eqnarray}
&& S_I= 
- 4e^2 \sum_{m, m'} [\Gamma_1^+(m')-\Gamma_1^-(m')-I/e] 
(\hat{\bf\Gamma}^{-1})_{m'm} 
        \nonumber \\
&& \hspace{0.3cm} \times [\Gamma_1^+(m-1) 
 \sigma_{st}(m-1) -\Gamma_1^-(m+1)\sigma_{st}(m+1)
\nonumber \\
&& \hspace{0.3cm}
-(I/e)\sigma_{st}(m)]
+ 2e^2 \sum_m \sigma_{st}(m) [\Gamma_1^+(m)+\Gamma_1^-(m)],
        \label{S_I}\end{eqnarray}
where $\hat{\bf \Gamma}$ is the three-diagonal matrix of the SET charge 
evolution, ${\bf \Gamma}_{km}=\sum_j [ \delta_{m,k-1}\Gamma_j^+(m)+
\delta_{m,k+1}\Gamma_j^-(m)-\delta_{m,k}[\Gamma_j^+(m)+\Gamma_j^-(m)] ]$.
Notice that Eq.\ (\ref{S_I}) can be easily used numerically even though 
the matrix $\hat{\bf \Gamma}$ is singular and therefore does not have a 
unique inverse; Eq.\ (\ref{S_I}) is constructed in a way that the 
non-uniqueness is not important, and therefore the standard algorithm for 
solving a linear system of equations with three-diagonal matrix 
can be readily used. 
(There is no problem of the matrix singularity at a finite frequency 
$\omega_s$, when the inverse of the matrix 
$\hat{\bf \Gamma}-\imat \omega_s \hat{\bf 1}$ should be calculated.) 
Actually, instead of using Eq.\ (\ref{S_I}), we have used a somewhat faster
algorithm for calculation of $S_I$, described in Section VII 
of Ref.\ \cite{Kor-94}.

\end{appendix}

\end{document}